\documentclass[a4paper]{article}

\usepackage{INTERSPEECH2020}
\usepackage{multirow}

\title{Domain Adaptation Using Class Similarity for Robust Speech Recognition}
\name{Han Zhu$^{1,2}$, Jiangjiang Zhao$^{3}$, Yuling Ren$^{3}$, Li Wang$^{1*\thanks{* Li Wang is the corresponding author.}}$, Pengyuan Zhang$^{1,2}$}
\address{
	$^1$Key Laboratory of Speech Acoustics and Content Understanding, Institute of Acoustics, China\\
	$^2$University of Chinese Academy of Sciences, China\\
	$^3$China Mobile Online Services Company Limited, China}
\email{\{zhuhan,wangli,zhangpengyuan\}@hccl.ioa.ac.cn, \{zhaojiangjiang,renyuling\}@cmos.chinamobile.com }

\begin{document}

\maketitle
\begin{abstract}
When only limited target domain data is available, domain adaptation could be used to promote performance of deep neural network (DNN) acoustic model by leveraging well-trained source model and target domain data. However, suffering from domain mismatch and data sparsity, domain adaptation is very challenging. This paper proposes a novel adaptation method for DNN acoustic model using class similarity. Since the output distribution of DNN model contains the knowledge of similarity among classes, which is applicable to both source and target domain, it could be transferred from source to target model for the performance improvement. In our approach, we first compute the frame level posterior probabilities of source samples using source model. Then, for each class, probabilities of this class are used to compute a mean vector, which we refer to as mean soft labels. During adaptation, these mean soft labels are used in a regularization term to train the target model. Experiments showed that our approach outperforms fine-tuning using one-hot labels on both accent and noise adaptation task, especially when source and target domain are highly mismatched.

\end{abstract}

\section{Introduction}\label{sec:Introduction}
Deep neural networks have significantly improved the state of the art for automatic speech recognition (ASR) \cite{dahl2012context}. However, this impressive gain in performance comes only when massive amounts of labeled data are available. For many scenarios, these training data are either difficult to collect, or expensive to annotate. Therefore, to address the scarcity of labeled data in target domains, we could leverage the massively available labeled data in related source ones. However, when domain mismatch occurs, such as speaker \cite{li2010comparison}, accent \cite{zhu2019multi} and environment \cite{li2017large} mismatch, the difference between data distributions of source and target domain increase the difficulty to use source data. Domain adaptation is often used to tackle this problem.

One of the major issues to be solved in domain adaptation is the data sparsity in target domain. With extremely sparse target domain data, overfitting problem is often encountered.

The direct cause of overfitting is too much parameters to be tuned during adaptation. Thus, the most intuitive approach is to constrain parameters to be trained. Selective fine-tuning \cite{sim2018domain} tries to find optimal number of layers to be tuned. Trainable parameters could also be reduced using singular value decomposition \cite{xue2014singular}. Another kind of approaches are to introduce new parameters into the neural network and only adapt these parameters during adaptation, such as linear transformation layers \cite{li2010comparison} \cite{gemello2006adaptation} and learning hidden unit contributions (LHUC) \cite{swietojanski2014learning}. The third kind of approaches are to enhance the input feature with domain discriminative vector, such as speaker code \cite{abdel2013fast} and i-vector \cite{saon2013speaker}. Furthermore, regularization based approaches are also widely used to handle overfitting, such as L2 regularization \cite{liao2013speaker} and KL-Divergence regularization \cite{yu2013kl}.  Adding these regularization term to the training criterion could avoid the model being driven too far from the original one. 

Overfitting problem could be alleviated when the amount of target domain data is above a specific threshold. However, this amount of data may still not be enough to adapt the source model to the target domain well. Soft labels are often beneficial to domain adaptation because they contain the similarity structure among predicted classes, which one-hot labels don't have.

Teacher-student (T/S) \cite{li2017large} \cite{meng2019conditional} learning proposed to promote domain adaptation performance using soft labels as well as parallel data consisting of pairs of samples from the source and target domain. Firstly, the source data is used to train the source model, which is used as teacher model. During adaptation, the teacher model computes posterior probabilities (soft labels) over source samples and the student model is trained on these soft labels and the parallel target samples.

However, parallel data of some domains is hard to be generated through simulation, e.\,g., accented speech. Knowledge distillation based domain adaptation \cite{asami2017domain} could be used to overcome this difficulty. In this approach, knowledge distillation \cite{hinton2015distilling} \cite{li2014learning} is combined with the T/S learning. During adaptation, target samples are fed into source model to get soft labels, which are used as an auxiliary task for the target model. An inevitable problem of this approach is that due to domain mismatch (i.\,e., using source model to compute over target samples), the computed posterior probabilities (soft labels) could be extremely wrong.

In order to alleviate this issue, inspired by \cite{tzeng2015simultaneous}, we propose to compute posterior probabilities using source model over all source samples. Then, for each class, we compute mean vectors of the corresponding probabilities generated by samples of this class. These vectors are referred to as mean soft labels, which are used instead of soft labels during adaptation. Experiments showed that our approach consistently outperform fine-tuning using one-hot labels on both accent adaptation and noise adaptation task. When source and target domain are highly mismatched, this approach further outperform KL-Divergence regularization and knowledge distillation based adaptation.

The rest of this paper is organized as follows, in Section \ref{sec:Related}, we describe some related work. In Section \ref{sec:class_similarity}, our approach is described in detail. Then, we evaluate our proposed approaches and compare them with other adaptation methods in Section \ref{sec:Experiments}, and conclude the study in Section \ref{sec:Conclusions}.

\section{Related work}\label{sec:Related}

As discussed in Section \ref{sec:Introduction}, there are many approaches proposed to improve domain adaptation performance. Among them, similar with our approach, teacher-student learning and knowledge distillation based adaptation achieved this goal by transferring similarity structure of predicted classes from the source to target domain. 

\subsection{Teacher-student learning}

In teacher-student (T/S) learning, the posterior probabilities generate by the source model can be used instead of one-hot labels to train the target model. In this approach, no labeled data is required. A corpus of unlabeled parallel data, which consisting of pairs of samples from source and the desired target domain, could be used instead.

In T/S learning, the source model $\theta_{S}$ trained on source data $\mathbf{X}^{S}=\left\{\mathbf{x}_{1}^{S}, \ldots, \mathbf{x}_{N}^{S}\right\}$ is used as teacher model and the target model $\theta_{T}$ trained on target data $\mathbf{X}^{T}=\left\{\mathbf{x}_{1}^{T}, \ldots, \mathbf{x}_{N}^{T}\right\}$ is used as student model. $N$ is the number of samples. The goal is to learn a student model that can accurately predict the class labels for its input sequence using the knowledge transferred from the teacher model.
To ensure effective knowledge transfer, the input sample sequences $\mathbf{X}^{T}$ and $\mathbf{X}^{S}$ need to be in parallel with each other, i.\,e., each pair of train samples $\mathbf{x}_{i}^{T}$ and $\mathbf{x}_{i}^{S}$ have the same class label $c_{i} \in\left\{1,2, \ldots, D_{C}\right\}$. $D_{C}$ is the amount of classes. 

To achieve that, we need to minimize the KL-divergence between the teacher and student output distributions $p\left(c | \mathbf{x}_{i}^{S} ; \theta_{S}\right)$ and $p\left(c | \mathbf{x}_{i}^{T} ; \theta_{T}\right)$, i.e, 
$ KL\left[p\left(c | \mathbf{x}_{i}^{S} ; \theta_{S}\right) \| p\left(c | \mathbf{x}_{i}^{T} ; \theta_{T}\right)\right]  $. And it is equivalent to minimizing the following loss function:
\begin{equation} \label{equ:kld}L_{\text{TS}}\left(\theta_{S}\right)=-\frac{1}{N} \sum_{i=1}^{N} \sum_{c=1}^{D_{C}} p\left(c | \mathbf{x}_{i}^{S} ; \theta_{S}\right) \log p\left(c | \mathbf{x}_{i}^{T} ; \theta_{T}\right)\end{equation}

\subsection{Knowledge distillation based adaptation}

Even though generating certain kinds of parallel data is convenient in many important scenarios, e.\,g., noisy or reverberant speech, there are still many important scenarios where parallel data is very difficult to generate, e.\,g., accented speech.

Thus, an adaptation approach using knowledge distillation, in which no parallel data is required, could be used instead. In this approach, the soft labels are computed using source model over target samples, which are used in regularization term to train the target model together with conventional one-hot labels.

The regularization term is also referred to as soft loss, which could be formulated as:

\begin{equation} \label{equ:soft_loss} L_{\text{soft}}\left(\theta_{T}\right)=-\frac{1}{N} \sum_{i=1}^{N} \sum_{c=1}^{D_{C}} q\left(c | \mathbf{x}_{i}^{T} ; \theta_{S}\right) \log q\left(c | \mathbf{x}_{i}^{T} ; \theta_{T}\right)\end{equation}
where $q\left(c | \mathbf{x}_{i}^{T} ; \theta_{S}\right)$ is the soft label computed using source model $\theta_{S}$. And $q\left(c | \mathbf{x}_{i}^{T} ; \theta_{T}\right)$ is the DNN output of target model $\theta_{T}$. $q\left(c | \mathbf{x}_{i}^{T} ; \theta\right)$ is the tempered softmax probability of the $c$-th class, which is computed over target samples $\mathbf{x}_{i}^{T}$ as follows:
\begin{equation}q\left(c | \mathbf{x}_{i}^{T} ; \theta\right)=\frac{\exp \left(z_{c}(\mathbf{x}_{i}^{T}; \theta) / T\right)}{\sum_{j=1}^{D_{C}} \exp \left(z_{j}(\mathbf{x}_{i}^{T}; \theta) / T\right)}\end{equation}
where $z_i(x; \theta)$ is the output of the i-th class of the model before softmax operation and $T$ is the temperature. Higher value for $T$ produces a softer probability distribution over classes \cite{hinton2015distilling}.

The cross entropy loss using one-hot labels is also referred to as hard loss, which is formulated as:
\begin{equation} \label{equ:hard_loss} L_{\text{hard}}\left(\theta_{T}\right)=-\frac{1}{N} \sum_{i=1}^{N} \sum_{c=1}^{D_{C}} y_{c} \log p\left(c | \mathbf{x}_{i}^{T} ; \theta_{T}\right)\end{equation}
where $y_c$ is the $c$-th dimension of the $D_{C}$ dimensional one-hot label.

Thus, the total loss function is as follows:
\begin{equation}\label{equ:distillation_loss} L_{\text{distillation}}= L_{\text{hard}}\left(\theta_{T}\right)+\rho L_{\text{soft}}\left(\theta_{T}\right)\end{equation}
where $\rho$ is the weight of the soft loss. Note that when $T = 1$, it is equivalent to KL-Divergence regularization.

During backpropagation, $T^2$
should be multiplied to the gradient of the soft loss. This ensures that the relative contributions of the hard and soft labels remain roughly unchanged when the temperature is changed.

\section{Adaptation using class similarity}\label{sec:class_similarity}

Knowledge distillation based adaptation is easy to implement for the reason that it wouldn't need parallel data. However, in this approach, as shown in (\ref{equ:soft_loss}), tempered probability $q\left(c | \mathbf{x}_{i}^{T} ; \theta_{S}\right)$ (i.\,e., the soft label) is computed using \emph{source} model $\theta_{S}$ over \emph{target} data $\mathbf{X}^{T}=\left\{\mathbf{x}_{1}^{T}, \ldots, \mathbf{x}_{N}^{T}\right\}$, which would lead to the domain mismatch. And the prediction accuracy of posterior probabilities would be significantly decreased.

We notice that in both teacher-student and knowledge distillation based adaptation, the key idea is to use soft labels. The soft label could improve performance by transferring relative similarity structure of predicted classes during adaptation \cite{asami2017domain}. 

In previous approach, each sample would produce a soft label. But we can't guarantee every sample could be predicted correctly. Moreover, the computation for each sample is not necessary since similarity structure among classes could be viewed as an inherent attribute, i.\,e., for each class, its relationship with other classes could be roughly determined without consideration of specific sample. Thus, each class could use a constant soft label to represent this structure.

\begin{figure*}[t]
	\centering
	\includegraphics[width=\linewidth]{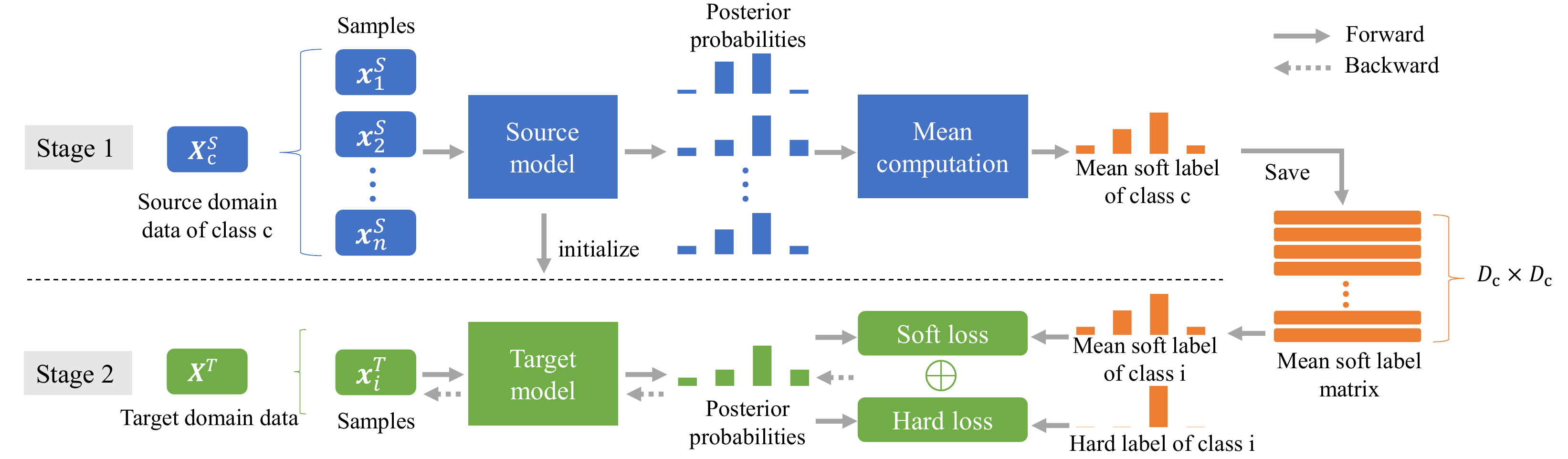}
	\caption{An overview of the proposed adaptation approach using mean soft label. In stage 1, for each class, we first compute the frame level posterior probabilities of source samples using source model. Then, the mean vector of  probabilities of each class is computed and saved. In stage 2, mean soft labels are used as a regularization term to train the target model. The regularization term and the conventional loss term are referred to as soft loss and hard loss respectively. }
	\label{fig:mean_soft_label}
\end{figure*}

To show why this is reasonable, we illustrate it as follows. Note that the predicted class of DNN acoustic is considered as phoneme instead of senones for simplicity. Take phoneme /ae/ as an example, its pronunciation is more similar to /e/ than /i/ no matter in the source or the target domain. Thus, the desired soft label of class /ae/ should have higher value on /e/ than /i/, which explicitly reflect the similarity structure among classes. Since source model trained on relatively more sufficient data, it could learn this similarity structure better than the target model. Thus, transferring this similarity structure from source domain is expected to improve performance on target domain.

Our approach is illustrated in Figure~\ref{fig:mean_soft_label}. To obtain the desired constant soft label, we first compute the tempered softmax probabilities $q\left(c | \mathbf{x}_{i}^{S} ; \theta_{S}\right)$ using source model $\theta_{S}$ over all source samples of the $c$-th class $\mathbf{X}^{S}_{c}$. Then we compute the mean value over all the $c$-th class probabilities to get the $c$-th mean soft label $l_{c}$ as follow:

\begin{equation}\label{equ:mean_soft_label} l_{c}= \frac{1}{N_{c}}\sum_{i=1}^{N_{c}}q\left(c | \mathbf{x}_{i}^{S} ; \theta_{S}\right), \mathbf{x}_{i}^{S} \in \mathbf{X}^{S}_{c}\end{equation}
where $N_{c}$ is the number of samples in $\mathbf{X}^{S}_{c}$. $l_{c}$ is used as the constant label of $c$-th class since it statistically encodes the similarity structure among predicted classes through the mean computation operation. 

During adaptation, mean soft labels are selected using one-hot label via the table look-up operation. Since mean soft labels are computed before adaptation, the computation cost during adaptation is also decreased compared with KL-Divergence regularization and knowledge distillation based approaches

The hard loss is same with equation (\ref{equ:hard_loss}) and the soft loss is computed as follows:

\begin{equation} \label{equ:mean_soft_loss} L_{\text{soft}}\left(\theta_{T}\right)=-\frac{1}{N} \sum_{i=1}^{N} \sum_{c=1}^{D_{C}} l_{c} \log q\left(c | \mathbf{x}_{i}^{T} ; \theta_{T}\right)\end{equation}

Finally, the total loss of our approach is:

\begin{equation}\label{equ:mean_soft_total_loss} L_{\text{cs}}= L_{\text{hard}}\left(\theta_{T}\right)+\rho L_{\text{soft}}\left(\theta_{T}\right)\end{equation}
where $\rho$ is the weight of the soft loss. Note that when $\rho$ is set to $\infty$, the target model is only trained with soft loss.

\section{Experiments and Results}\label{sec:Experiments}

To analyze the effectiveness of our method, two sets of experiments are conducted: accent adaptation and noise adaptation. 

\subsection{Accent adaptation}
For accent robustness experiments, we use the Common Voice corpus \cite{ardila2020common} from Mozilla. United States English (US) are used as source domain. In terms of the target domain, we choose three well represent accents: England English (EN), Australian English (AU), and South Asian English (IN). US training set contains 30000 sentences, while training sets of other accents contain 2000 sentences. Validation sets and Testing sets of all accents include 1000 sentences. 

To get senone-level forced alignment, a GMM-HMM system is trained on source domain data (i.\,e., US speech data). The GMM-HMM system is trained using Kaldi \cite{povey2011kaldi}. Forced alignment of both source and target data are all generated using this source domain GMM-HMM system. 39-dimensional MFCC feature (13 static+$\Delta$+$\Delta\Delta$) is used in all experiments. The features are fed as the input of the acoustic model after global mean and variance normalization. 

The DNN acoustic models are trained using PyTorch-Kaldi \cite{ravanelli2019pytorch}.The source model is used to initialize the target model before adaption. The source and target model have the same structure consisting of four 550 dimensional bidirectional gated recurrent unit (GRU) layers. The output layer has 3280 units. Recurrent dropout was used as a regularization technique \cite{moon2015rnndrop}. Batch normalization was adopted for feed-forward connections only \cite{ravanelli2017improving}. The optimization was done using the RMSprop algorithm \cite{tieleman2012lecture}. The performance on the development set was monitored after each epoch and the learning rate was halved when the relative performance improvement went below 0.1\%. Initial learning rate is set to 0.004. Training is stopped when learning rate is less than 10\% of the initial learning rate. The model achieving the best validation accuracy is used for the test.

The source GRU acoustic model achieves 7.61\% WER on source domain testing set. Performance on target domain is shown in Table \ref{tab:Result}.

\begin{table*}[t]
	\caption{Domain adaptation performance for accent/noise adaptation in terms of WER (Word Error Rate). The best results in each column are highlighted in bold. \textbf{Source/target only} denotes the model trained on only source/target data. \textbf{Source + target} model trained on both source and target data. \textbf{Fine-tune}, \textbf{KLD}, \textbf{distillation} and \textbf{mean soft label} denote fine-tuning using one-hot labels, KL-Divergence regularization, knowledge distillation based adaptation and adaptation using mean soft labels respectively.}
	\label{tab:Result}
	\centering
	\begin{tabular}{@{}c|cc|ccc|ccccc@{}}
		\toprule
		&   &   & \multicolumn{3}{c|}{\textbf{Accent Adaptation}}                  & \multicolumn{5}{c}{\textbf{Noise Adaptation}}                                     \\ \hline
		& \bm{$T$} & \bm{$\rho$} & \textbf{EN}    & \textbf{AU}   & \textbf{IN}      & \textbf{BUS}   & \textbf{CAF}   & \textbf{PED}   & \textbf{STR}   & \textbf{AVG}   \\
		\hline
		source only                      & -                  & -   & 13.73         & 16.38          & 30.65          & 83.23          & 68.45          & 67.36          & 43.98          & 65.42          \\
		\hline
		target only                      & -                  & -   & 15.41         & 16.88          & 15.35          & 46.09          & 38.53          & 32.96          & 22.16          & 34.70          \\
		\hline
		source + target                     & -                  & -   & 10.36 &12.75   & 17.15             & 61.61 & 41.23 &   36.69    & 22.18  & 39.99          \\
		\hline
		fine-tune                        & -                  & -   & 8.94          & 11.29          & 12.31          & 42.84          & 33.39          & 28.19          & 18.10          & 30.37          \\
		\hline
		\multirow{4}{*}{KLD}             & \multirow{4}{*}{1} & 0.1 & 8.69          & 11.20          & 11.55          & 54.18          & 36.11          & 30.50          & 22.59          & 35.45          \\
		&                    & 0.2 & 8.51          & 10.91          & 11.68          & 63.84          & 37.89          & 31.95          & 23.76          & 38.84          \\
		&                    & 0.5 & 8.55          & 11.26          & 12.40          & 73.72          & 43.18          & 37.37          & 28.04          & 44.98          \\
		&                    & 1   & 8.63          & 11.49          & 13.76          & 81.90          & 48.23          & 44.04          & 32.97          & 51.15          \\
		\hline
		\multirow{8}{*}{distillation}    & \multirow{4}{*}{2} & 0.1 & 8.55          & 10.69          & 11.61          & 60.77          & 35.43          & 30.11          & 22.72          & 36.76          \\
		&                    & 0.2 & \textbf{8.43} & 10.85          & 11.68          & 65.69          & 37.42          & 31.65          & 23.80          & 39.08          \\
		&                    & 0.5 & 8.60          & 11.10          & 12.21          & 74.91          & 40.53          & 36.65          & 27.33          & 44.22          \\
		&                    & 1   & 8.94          & 11.88          & 13.45          & 81.92          & 47.66          & 43.35          & 31.43          & 50.44          \\
		\cline{2-11}
		& \multirow{4}{*}{5} & 0.1 & 8.62          & \textbf{10.58} & 11.50          & 54.45          & 34.23          & 29.38          & 20.55          & 34.23          \\
		&                    & 0.2 & 8.53          & \textbf{10.58} & 11.77          & 60.42          & 35.43          & 29.39          & 21.51          & 36.18          \\
		&                    & 0.5 & 8.54          & 10.61          & 12.01          & 69.22          & 38.65          & 33.56          & 24.48          & 40.89          \\
		&                    & 1   & 8.53          & 11.38          & 12.93          & 78.08          & 43.82          & 39.35          & 27.79          & 46.62          \\
		\hline
		\multirow{5}{*}{mean soft label} & \multirow{5}{*}{1} & 0.1 & 8.84          & 10.82          & 12.18          & 42.00          & 33.20          & 28.34          & 18.27          & 30.21          \\
		&                    & 0.2 & 8.83          & 10.69          & \textbf{10.94} & 40.25          & 32.73          & 27.82          & 18.14          & 29.51          \\
		&                    & 0.5 & 8.73          & 10.62          & 10.95          & 40.27          & \textbf{32.21} & 28.02          & \textbf{17.97} & \textbf{29.40} \\
		&                    & 1   & 8.66          & 10.61          & 11.91          & \textbf{40.17} & 32.64          & \textbf{27.76} & 18.20          & 29.47   \\
		&                    & $\infty$   & 9.49          & 10.69          & 11.04          & 42.19 & 33.41 &   28.25  &   18.74  & 30.40 
		\\ \bottomrule
	\end{tabular}
\end{table*}

In comparison with our approach using mean soft labels, we conducted fine-tuning using one-hot labels, KL-Divergence regularization and knowledge distillation based adaptation experiments. The main hyperparameters to be tuned are temperature $T$ and the soft loss weight $\rho$. We conducted experiments with $T = [1, 2, 5]$ and $\rho = [0.1, 0.2, 0.5, 1]$. For adaptation using mean soft label, high temperature has no significant performance improvement in our experiments. Thus we only report results with $T = 1$. $\rho = \infty$ experiments are also conducted to analyze the performance when only the soft loss is used. Results are reported in the accent adaptation part of the Table \ref{tab:Result}.

We could observe that adaptation using mean soft labels consistently outperform fine-tuning on all three accents and further outperform KL-Divergence and knowledge distillation based adaptation on IN. However, in terms of EN and AU, knowledge distillation based adaptation performs best. 

In order to analyze results, we further test the frame level accuracy of the source model on EN, AU and IN testing sets, which are are 62.8\%, 60.8\% and 52.0\% respectively, i.\,e., the source model performs better on EN and AU in terms of both frame level accuracy and WER.

This result is consistent with intuition. Note that different samples of the same class could also have different soft labels because the every sample have its unique acoustic feature. However, adaptation using mean soft labels can't take such variability into account. In contrast, knowledge distillation based adaptation could assign different soft labels to different samples of the same class. This is an advantage of knowledge distillation based adaptation approach. When domain mismatch is critical and source model performs poorly on target domain, the performance gain due to this advantage is less than the degradation coming from the domain mismatch. In this case, adaptation using mean soft labels could outperform knowledge distillation based approach,  which is the case for accent IN. However, in terms of accent EN and AU, errors caused by domain mismatch is relatively less so that the knowledge distillation based approach performs better.

Thus, we argue that the proposed approach should perform better on the conditions where source and target domain are highly mismatched. Experiments of noise adaptation could be further conducted to verify this hypothesis. 

\subsection{Noise adaptation}

For noise adaptation experiments, we use the CHiME-3 \cite{barker2015third} dataset. The CHiME-3
dataset incorporates Wall Street Journal (WSJ) corpus \cite{paul1992design} sentences spoken in challenging noisy environments, recorded using a 6-channel tablet. The clean data from WSJ0 SI-84 dataset is used as source domain. For the target domain, We use the real single-channel far-field noisy speech (the 5th microphone channel). The dataset splitting of training, validation and testing set is the same as the official dataset. The testing set consists of real data recorded with 4 locations, i.\,e., on the bus (BUS), cafe (CAF), pedestrian area (PED), and street junction (STR).

The experiment setup is similar with accent adaptation. The differences between them are described below. The force alignment of clean and noisy data is generated by two different GMM-HMM systems trained on corresponding training sets. The output layer of the GRU model has 2008 output units.

The clean CHiME-3 GRU acoustic model achieves 2.26\% and 65.42\% WER on clean and noisy testing sets respectively. And the frame level frame accuracy on noisy testing set is only 17.0\% , i.\,e., the source and target domain are highly mismatch. The adaptation performance results are reported in the noise adaptation part of the Table \ref{tab:Result}.

KL-Divergence regularization and knowledge distillation based adaptation both perform worse than fine-tuning, which is caused by the domain mismatch issue. In contrast, our approach using mean soft labels outperform fine-tuning with 0.97\% absolute WER reduction in terms of average performance.

These results are consistent with our hypothesis, i.\,e., adaptation using mean soft labels works well when source and target domain are highly mismatched. Furthermore, in all experiments, using mean soft labels together with one-hot labels performs better than using soft labels alone ($\rho = \infty$).

\section{Conclusions}\label{sec:Conclusions}

In this paper, we proposed a novel adaptation technique for DNN acoustic model using mean soft labels. On the one hand, the class similarity structure of DNN targets is leveraged to prompt knowledge transferring. On the other hand, it alleviates the domain mismatch issue using mean soft labels which are computed using source model over source data.

Experiments showed that our approach outperforms fine-tuning using one-hot labels on both accent and noise adaptation task. When the source and target domain are highly mismatched, it further outperforms KL-Divergence regularization and knowledge distillation based adaptation.

\section{Acknowledgements}
This work is partially supported by the National Natural Science Foundation of China (Nos. 11590774,11590772,11590770)

\bibliographystyle{IEEEtran}

\bibliography{mybib}

\end{document}